\documentclass[11pt,a4paper]{article}
\usepackage[hyperref]{emnlp2020}
\pdfoutput=1
\usepackage{times}
\usepackage{latexsym}

\usepackage{microtype}

\aclfinalcopy 
\def\aclpaperid{2740} 

\usepackage[export]{adjustbox}

\usepackage{times}  
\usepackage{helvet} 
\usepackage{courier}  
\usepackage{url}  
\usepackage{graphicx} 
\usepackage{booktabs}       
\usepackage{amsfonts}       
\usepackage{nicefrac} 
\usepackage{latexsym}
\usepackage{multirow}
\usepackage{booktabs}
\usepackage{amsmath}
\usepackage{booktabs}
\usepackage{courier}
\usepackage{colortbl}
\usepackage[position=bottom]{subfig}
\usepackage{microtype}
\usepackage{array}
\usepackage{wrapfig}
\usepackage{color,xcolor}

\usepackage{alphalph}
\usepackage{caption}
\usepackage{makecell}

\title{Event-Driven Learning of Systematic Behaviours in Stock Markets}

\author{Xianchao Wu  \\
  {\rm NVIDIA} \\
  {\tt \normalsize xianchaow@nvidia.com, wuxianchao@gmail.com}}

\date{}

\begin{document}
\maketitle
\begin{abstract}

It is reported that financial news, especially financial events expressed in news, provide information to investors' long/short decisions and influence the movements of stock markets. Motivated by this, we leverage financial event streams to train a classification neural network that detects latent event-stock linkages and stock markets' systematic behaviours in the U.S. stock market. Our proposed pipeline includes (1) a combined event extraction method that utilizes Open Information Extraction and neural co-reference resolution, (2) a BERT/ALBERT enhanced representation of events, and (3) an extended hierarchical attention network that includes attentions on event, news and temporal levels. Our pipeline achieves significantly better accuracies and higher simulated annualized returns than state-of-the-art models when being applied to predicting Standard\&Poor 500, Dow Jones, Nasdaq indices and 10 individual stocks. 

\end{abstract}

\section{Introduction}

It is widely reported that financial news, especially financial events expressed in the news, influence the movements of stock markets \cite{ding-etal-2014-using, ding-etal-2015-ijcai, ding-etal-2016-knowledge,hu.wsdm2018, ding-etal-2019-event-representation, Huang2014InstitutionalTA, Glasserman2019TimeVI}. For example, one company's releasing of new products brings novel time-sensitive factors to the stock movement of that company and further to the whole market; changing of interest rates from the central bank of the country bring changes of currency fluidity invested in the stock market. Frequently, good news brings increased estimation of the future value of the target company and consequently a higher stock price tendency.

Warren Buffett said in his interview\footnote{\url{https://www.youtube.com/watch?v=Pqc56crs56s}} that he frequently spent 5 hours reading news and financial reports to manage his portfolios. Regarding the rich information included in daily published news articles, we are encouraged to read them not by ourselves but by employing deep learning algorithms to guide our investments. One solution is to automatically collect thousands of news published every day, extract financial events from the news, and rank the importance of the events to predict market behaviours. In particular, we are aiming at quantifying the \emph{latent relevance} of between financial event streams and target stocks' price volatilities. 

However, this is not a trivial task and there are a number of challenges. First, it is ambiguous to define good or bad news. For example, bad news for one company could be worse news to its downstream supply-chain partner companies yet good news to its competitors. It is time-consuming and untrackable to annotate news manually and train a sentiment analysis model on it, considering that the generalization of the model to novel financial events is fragile. Second, how shall we express the \emph{impact} of the news published in different days? Generally, news articles have their individual and accumulated influences to the investors. For example, billion-dollar mergers and acquisitions frequently bring bigger and longer impact than adding a new member to the board. Third, regarding that news articles are too long to be generalized for comparison, how to extract shorter, high-level summarized and complete financial events from news? There are factual and opinion-level events expressed in news and their appearance order matters for concluding the news. Finally, how to measure the similarities among financial events guided by historical stock movements? The latent-space representation of events empowers the generalization ability of a machine learning model for predicting event-volatility based on representing and projecting novel events into existing event representation space. 

In order to tackle these challenges, we propose a classification network for stock movement prediction. We first describe a combined event extraction method basing on Open Information Extraction \cite{fader-etal-2011-identifying} and neural co-reference resolution \cite{clark-manning-2016-improving, clark-manning-2016-deep} to keep the mined events to be compact and complete with meaning. Then, we learn the event representations and similarities among events under BERT/ALBERT \cite{devlin-etal-2019-bert,Lan2020ALBERT:} pretrained contextualized language models. Finally, we construct a hierarchical attention network (HAN) \cite{yang-etal-2016-hierarchical,hu.wsdm2018} that employs events, news, and historical days to organize the granularity of information for final multi-category stock movement prediction.

\section{Event Definition and Extraction}

\subsection{Event Definition and Classification}

Following \newcite{ding-etal-2014-using}, we define a financial event as a tuple alike $\langle a_1, p, a_2, [timestamp]\rangle$. Arguments $a_1$ and $a_2$ respectively act as subject and object. The predicate $p$ is the action that links $a_1$ and $a_2$. Publishing timestamp of the news is attached to each tuple, which is used to align events with the consequent stock movements. 

The major components in arguments $a_1$ and $a_2$ are named entities (such as names of person, company, and stock/index). The main components in predicates $p$ are verb (phrases) standing for actions performed of among the arguments. For example, ``\emph{the standard\&poor’s 500 index}, \emph{rose}, \emph{0.6 percent}''. The polarities of events are traditionally classified into positive, negative, and neutral events \cite{Huang2014InstitutionalTA}. In this paper, instead of explicitly assign polarity to each event, we ask the HAN model to tune the \emph{attentional weights} of the event sequences which are mixtures of various types of events expressed in news during a period.

From another point of view, events can be classified into \emph{objective evidence} and \emph{subjective opinions}. For example, ``\emph{equities and bonds}, \emph{were both in}, \emph{bear markets}'' is a real-world fact evidence. On the other hand, ``\emph{This}, \emph{is}, \emph{the best buying opportunity}'' is more alike a subjective opinion of someone's (e.g., journalists or analysts in the news) judgement. Our usage of event sequence is a mixture of evidence and opinions. The weights of events are learned simultaneously based on their contributions to the consequent days' stock movements.

\subsection{Event Extraction}

\begin{figure}
    \centering
    \includegraphics[width=7.5cm]{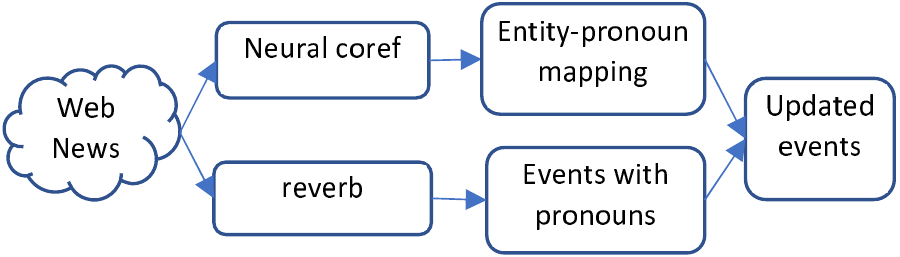}
    \caption{Applying reverb and neural-coref in parallel for event extracting.}
    \label{figure:event_extraction_pipeline}
\end{figure}

Our process for English financial event extraction is depicted in Figure \ref{figure:event_extraction_pipeline}. There are mainly two modules, the OpenIE reverb module\footnote{{\url{{https://github.com/knowitall/reverb}}}} \cite{fader-etal-2011-identifying} for raw event extraction and the neural coref module\footnote{{\url{https://github.com/huggingface/neuralcoref}}} \cite{clark-manning-2016-improving, clark-manning-2016-deep} for coreference resolution. These modules are executed in parallel and combined together to yield coreference-free events. There are two differences between our work and the former event-driven researches \cite{ding-etal-2014-using, ding-etal-2015-ijcai, ding-etal-2016-knowledge, ding-etal-2019-event-representation}: we additionally use neural-coref for entity linking to rewrite events and we ignore post-filtering such as restricted dependency relations among $a_1$, $a_2$ and $p$.

\begin{table} \footnotesize
    \centering
    \begin{tabular}{r|l}
    \hline
No. & Events (Original Format) \\
\hline\hline
1 &	Wednesday U.S. investment-grade corporate \\ 
 & bonds, have been heavily battered in, recent weeks \\
2 &	Wednesday U.S. investment-grade corporate \\
 & bonds, are, the best buying opportunity \\
3 &	This, is, the best buying opportunity \\
4 &	equities and bonds, were both in, bear markets \\
5 &	average now yield about 7 percent \\
6 &	That, 's almost as much as, junk bonds \\
7 &	Congress, will pass, a controversial \$ 700 billion \\
 & financial bailout package \\
8 &	the heels of the collapse of Lehman Brothers and \\
 & AIG, has triggered, a flight \\
9 &	people, have to sell for, one reason \\
10 & It, 's, an illiquid market \\
11 & an illiquid market, makes, matters \\
12 & Fuss, is vice chairman of, Boston-based \\
 & Loomis Sayles \\
13 & He/fuss, 's able to buy, long-maturing AA \\
14 & junk bonds, were yielding in, March 2007 \\
15 & vice chairman of Boston-based Loomis Sayles, \\
 & oversees more than, \$ 100 billion \\
\hline
    \end{tabular}
    \caption{Event examples extracted from one news.}
    \label{tab:event_extraction_example_table}
\end{table}

We select one article from Reuters\footnote{{\url{https://www.reuters.com/article/us-usa-bonds-loomissayles/loomis-sayles-fuss-sees-value-in-corporate-bonds-idUSTRE4906YT20081001}}} as an example. The extracted events are listed in Table \ref{tab:event_extraction_example_table}. These events reflect objective evidence (events 1, 4 to 10, 12 to 15) and subjective opinions (events 2, 3, 11). This news was published at 2008/10/01, right during the ``\emph{2008 Financial Crisis}''. It was an illiquid market with quite decreased trading volume of both stocks and bonds. Yet it was hard to judge whether or not it was the ``\emph{best}'' buying opportunity (Event 3), depending on how far the market trusts these opinions. With the usage of these evidence and opinions, we are hoping to learn latent connection of between these events and the next-day's (open-price) stock movement. 





\section{BERT-enhanced HANs with Three-Level Attentions}

\begin{figure}
    \centering
    \includegraphics[width=7.5cm]{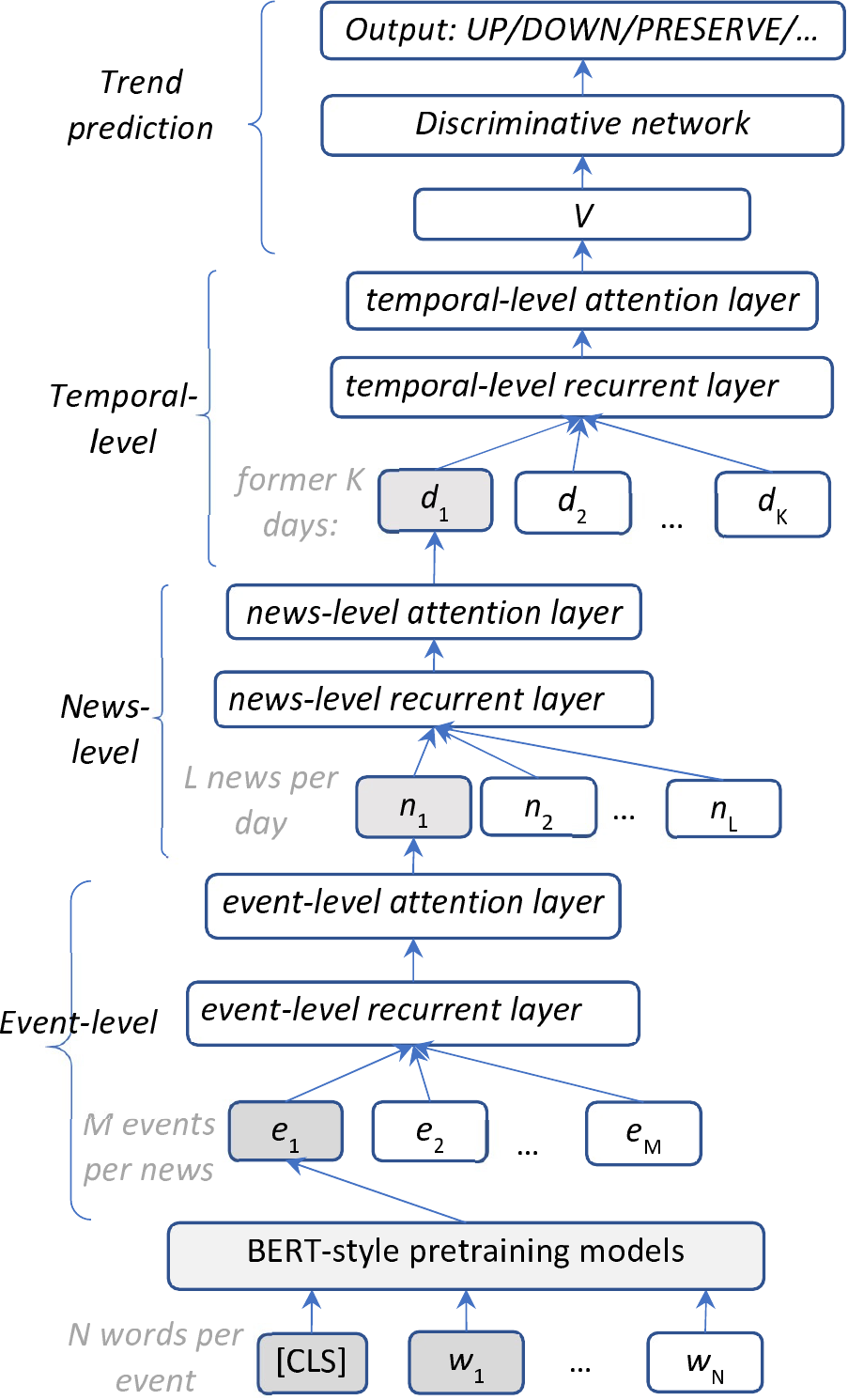}
    \caption{BERT-enhanced HAN with word, event, news and temporal level representations and attentions.}
    \label{figure:han}
\end{figure}

We propose two updates of the original HANs used in \cite{hu.wsdm2018} and \cite{yang-etal-2016-hierarchical}: inserting an event-layer between words and news levels, and representing events by pretrained contextualized language models such as BERT \cite{devlin-etal-2019-bert} instead of bidirectional gated recurrent unit (bi-GRU) \cite{cho-etal-2014-learning} plus attention networks.
The encoding of financial information in the units of events is inspired by \newcite{ding-etal-2014-using, ding-etal-2019-event-representation}. The difference is that, we replace the neural tensor network \cite{ding-etal-2014-using} by BERT. \newcite{hu.wsdm2018} directly used selected words in news as the initial layer in their HAN for stock movement prediction. The drawbacks that we find are (1) one news include hundreds to thousands words and it is difficult to select representative words from them to fit the final stock movement prediction task, and (2) too long news prevents the generalization ability of being embedded for financial information similarity computing.

Generally, our proposed network can be seen as a combination of events \cite{ding-etal-2014-using, ding-etal-2019-event-representation} represented by deep pretrained contextualized language models \cite{devlin-etal-2019-bert} inside a hierarchical attention mechanisms \cite{hu.wsdm2018, yang-etal-2016-hierarchical}. In Figure \ref{figure:han}, we assume that there are $N$ words in one financial event. We attach a classification token [CLS] at the beginning of an event and then execute BERT to obtain the representation tensor of the event. The output vector of [CLS] is taken as the representation of the event.

In the event-level network block, our target is to construct a \emph{recurrent} representation for the news by taking all its $M$ events into consideration. We first apply a bi-GRU to the vectors of events: 
\begin{eqnarray}
  \overrightarrow{h}_i & = & \overrightarrow{GRU}(e_i), i\in [1, M], \nonumber \\
  \overleftarrow{h}_i & = & \overleftarrow{GRU}(e_i), i\in [M, 1], \nonumber \\
  h_i & = & [\overrightarrow{h}^{\top}_i, \overleftarrow{h}^{\top}_i]^{\top}.\nonumber 
\end{eqnarray}
The result $h_i$ incorporates the contextual information of $M$ events. Through this way, we encode the event sequence of each news.

Considering that different events contribute unequally to the final representation of one news, we adopt the attention mechanism \cite{DBLP:journals/corr/BahdanauCB14,hu.wsdm2018} to aggregate the events weighted by an assigned attention value, in order to reward the event(s) that offer critical information: 
\begin{eqnarray}
  u_i & = & sigmoid(W_nh_i + b_n), \nonumber \\
  \beta_i & = & \frac{exp(\theta_ih_i)}{\sum_{i}{exp(\theta_ih_i)}}, \nonumber \\
  n_i & = & \sum_{i=1}^{M}{\beta_ih_i}. \nonumber
\end{eqnarray}
We first estimate attention values by feeding $h_i$ through a one-layer full-connection linear neural network followed by a sigmoid function to obtain the event-level attention value $u_i$. Then, we calculate its normalized attention weight $\beta_i$ through a softmax function. Finally, we calculate the news vector $n_i$ as a weighted sum of each event vector. A parameter $\theta_i$ is attached to each event in the softmax layer, indicating in general which event is more representative. Through this way, we construct the event-level representation and attention for each news. 

We reuse these recurrent and attention layers to news-level block. Suppose that there are maximum $L$ news in one day's financial news collection. We construct news-level recurrent and attention computing for one day's news. Here, bi-GRU is employed to capture the ``contextual'' relations among news sorted by their published timestamps. Also, attention mechanism is employed to capture which news contributes more to the representation of that day's vector $d_i$. Again, $d_i$ is an attention weighted sum of the news vectors. 

In the temporal level, we suppose that there are $K$ historical days for tomorrow's stock movement prediction. The news published at different dates contribute differently to the final stock trend. We the third time use bi-GRU to capture the latent dependencies of day vectors and attention mechanism to weight the day vectors. The final vector $V$ is a weighted sum of $d_1$ to $d_K$. 

The final discriminative network is a standard multi-layer perceptron (MLP) that takes $V$ as the input and produces the multi-category classification of the stock movement. Generally, the prediction is \emph{explainable} by listing the most valuable \emph{events} in the most trustable \emph{news} published in those important historical \emph{days} (Table \ref{table_c4_7_5c_eventlist}).


\section{Experiments}

\subsection{S\&P 500, Dow Jones and Nasdaq Indices}

We select Standard\&Poor (S\&P) 500, Dow Jones and Nasdaq indices\footnote{\url{https://finance.yahoo.com/}} to disclose how far financial news information can bring impacts to the \emph{systematic behaviours} in stock markets. Following researches on stock movement prediction \cite{hu.wsdm2018,ding-etal-2019-event-representation}, we define it as a multi-category classification problem.  

For a given date $t$ and a given stock $s$ (i.e., an individual stock or an index), its daily return $r(s,t)$ (or, rise percent) is computed by:
\begin{eqnarray}
    r(s,t)=\frac{open\_price(s,t)-open\_price(s,t-1)}{open\_price(s,t-1)} \nonumber
\end{eqnarray}

\begin{figure}
    \centering
    \includegraphics[width=7.6cm]{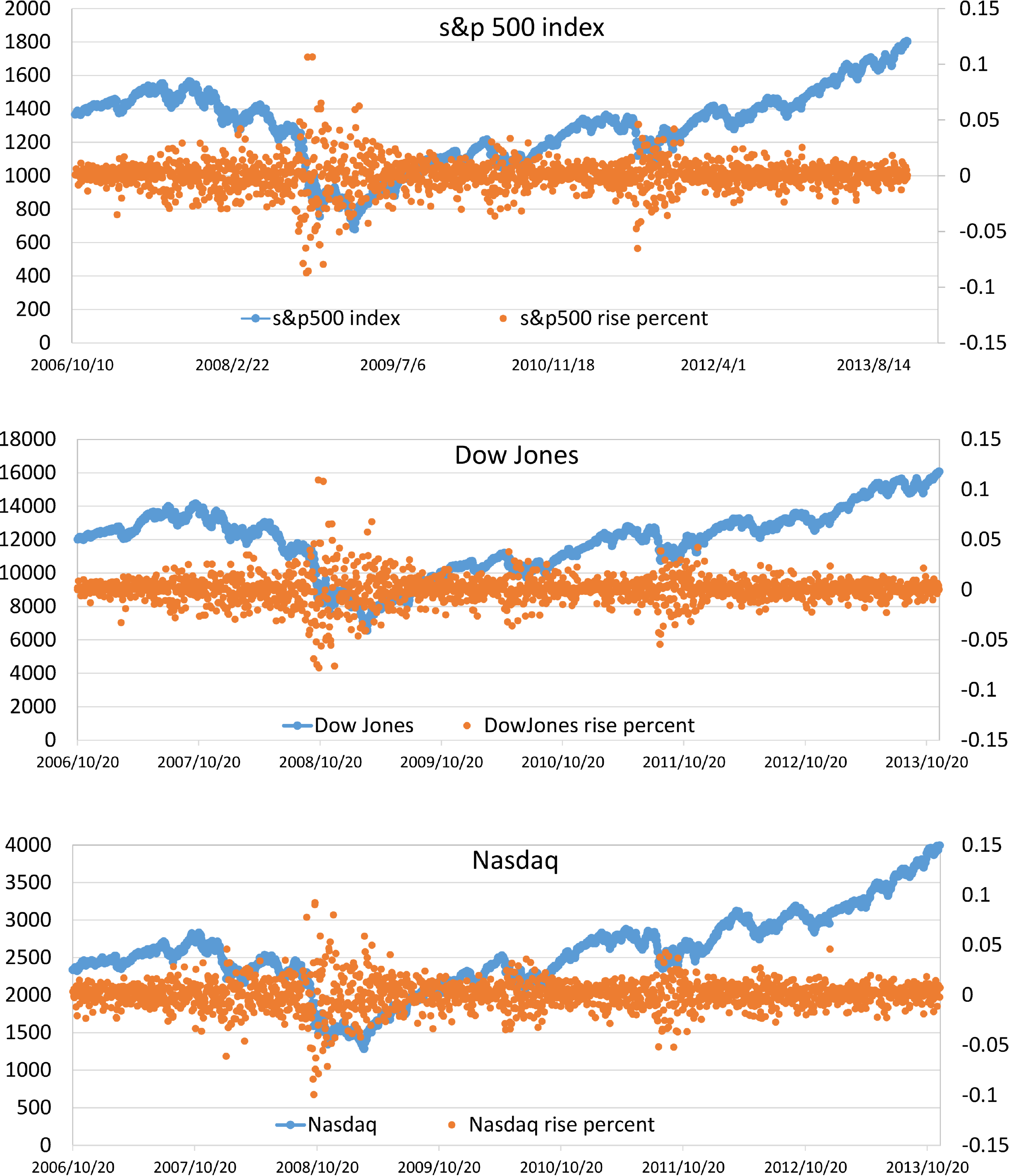}
    \caption{S\&P 500, Dow Jones, and Nasdaq indices (open prices) and their daily returns during the experiment period of from 2006/10/20 to 2013/11/27.}
    \label{figure:sp500_values}
\end{figure}

Date $t-1$ here refers to target stock's right-former market-opening date before date $t$. These three indices and their related daily returns ($r$) are depicted in Figure \ref{figure:sp500_values}.
Intuitively observing, despite their absolute values, the curves of these indices' open prices are quite similar. Thus, it is reasonable to argue that there do exist (latent) driving factors. Specifically, when we simply compare the UP ($r>0$) or DOWN ($r<0$) movements, S\&P 500 respectively shares 87.8\% and 75.4\% days with Dow Jones and Nasdaq of identical movements, and Dow Jones shares 69.2\% identical days with Nasdaq. 

For example, 2008 Financial Crisis caused a significant drop (as much as 50\%) of all the three indices during the whole 2008. After that, the indices recovered and continued increasing generally from 2009 to 2013. We separate this period into three subsets, the training set (2006/10/20 $\sim$ 2012/06/18), the validation set (2012/06/19 $\sim$ 2013/02/21), and the testing set (2013/02/22 $\sim$ 2013/11/27), exactly same with \cite{ding-etal-2014-using,ding-etal-2019-event-representation} for direct comparison\footnote{Note that the start date and end date are slightly different: \cite{ding-etal-2014-using} used 2006/10/02 and 2013/11/21 respectively. We made this shift by the number of available news are mostly zero outside these periods.}. The whole 2008 Financial Crisis period is included in the training set. We focus on out-of-sample testing to figure out if the weights of the events happened during the train period is explainable enough for the future movements of the indices.

For qualitatively evaluating the strengths of events, we define five categories, UP+, UP, PRESERVE, DOWN and DOWN-, representing the significant rising, rising, steady, dropping, and significant dropping compared with the former open-market date. In addition to align with \cite{ding-etal-2019-event-representation} and \cite{hu.wsdm2018}, we also report results on 2-category UP($r>0$)/DOWN($r<0$) and 3-category UP/PRESERVE/DOWN predictions.

There are 1,786 samples in our observation period and each sample contains more than one news. In order to balance the number of samples in each category, following \cite{hu.wsdm2018}, we respectively split this sample set equally into five and three subsets by setting four and two thresholds, for 5-category and 3-category classification. For example, for the 3-category case of S\&P 500 index, the lower/higher thresholds are -0.23\% and 0.38\%, yielding 601/609/576 DOWN/UP/PRESERVE samples.

\begin{table} \footnotesize
    \centering
    \begin{tabular}{l|rrr}
    \hline
Bloomberg & Train & Validation & Test \\
\hline
\# news days & 1,418 & 248 & 278 \\ 
\# news & 277,272 & 83,569 & 87,554 \\ 
\# news per day & 195.5 & 337.0 & 314.9 \\ 
\# sentences & 4,626,102 & 1,772,501 & 1,980,442 \\ 
\# words & 131,503,427 & 50,319,161 & 56,086,226 \\ 
\# events & 2,902,477 & 2,027,982 & 2,276,708 \\ 
\# events$/$news & 10.5 & 24.3 & 26.0 \\ 
\# events$/$day & 2046.9 & 8177.3 & 8189.6 \\ 
\# words in & & & \\ 
\ \ \ events & 21,727,751 & 14,072,051 & 17,052,547 \\ 
\# words$/$event & 7.5 & 6.9 & 7.5 \\ 
\hline \hline
Reuters & Train & Validation & Test \\ 
\hline
\# news days & 754 & 90 & 103 \\ 
\# news & 28,903 & 4,211 & 4,444 \\ 
\# news$/$day & 38.3 & 46.8 & 43.1 \\ 
\# sentences & 589,072 & 85,482 & 90,609 \\ 
\# words & 16,595,909 & 2,478,750 & 2,615,477 \\ 
\# events & 2,549,109 & 401,615 & 383,851 \\ 
\# events$/$news & 88.2 & 95.4 & 86.4 \\ 
\# events$/$day & 3380.8 & 4462.4 & 3726.7 \\ 
\# words in & & & \\
\ \ \ events & 19,571,840 & 3,100,215 & 2,966,719 \\ 
\# words$/$event & 7.7 & 7.7 & 7.7 \\ 
\hline\hline
\# samples & 1,423 & 168 & 195 \\
\hline
    \end{tabular}
    \caption{Statistical information (news and events related) for train, validation and test sets.}
    \label{tab:num_news_events}
\end{table}

\subsection{Bloomberg and Reuters News}

Following \cite{ding-etal-2015-ijcai, ding-etal-2016-knowledge, ding-etal-2019-event-representation}, we utilize Bloomberg and Reuters financial news\footnote{News data are available at \url{https://drive.google.com /drive/folders/0B3C8GEFwm08QY3AySmE2Z1daaUE}} (2006/10 to 2013/11) for extracting financial events that are related to the U.S. stock market. 

Table \ref{tab:num_news_events} lists detailed statistical information of the events that we extracted from Bloomberg and Reuters. The number of events in total is more than 10 million, significantly (around 28 times) larger than the 366K events used in \cite{ding-etal-2014-using, ding-etal-2016-knowledge, ding-etal-2019-event-representation}. We will show that these large-scale mined events are essential for training the BERT+HAN and for achieving significant better prediction accuracies (Section \ref{subsec:res_baselines}).

Bloomberg (ranges from 195.5 to 337.0 news articles per day) has much more news per day than Reuters (ranges from 38.3 to 46.8 news articles per day) during the three observation periods. Also, there are more missing days without any news in the Reuters dataset. In terms of events in the Bloomberg set, the three datasets (train, validation, and test) all have more than 2 million events and daily event number is in a range of from 2K (train set) to more than 8K (validation and test sets). These too long sequences prevent a direct usage of BERT style pretraining models which directly make a prediction based on [CLS]'s vector, due to the GPU memory limitation and computational difficulty of multi-head self-attention \cite{NIPS2017_7181_transformer} on 8K events in which each event further has averagely $>$7 words. 


\subsection{Experiment Setups}

In our HAN, we set historical days $K$=10, maximum news per day $L$=500, maximum events per news $M$=100 and maximum words per event $N$=20, the dimensions of hidden states in recurrent networks and attention vectors are set to be 1,024. We implement our BERT-HAN under Huggingface's transformers\footnote{\url{https://github.com/huggingface/transformers}} written in PyTorch. We specially selected BERT \cite{devlin-etal-2019-bert} pretrained model of ``bert-large-uncased'' and ALBERT \cite{Lan2020ALBERT:} pretrained model of ``albert-xxlarge-v2''\footnote{\url{https://github.com/google-research/ALBERT}}. Categorical cross-entropy loss is optimized by the Adam algorithm with weight decay \cite{Kingma2015AdamAM,loshchilov2018decoupled}. We run experiments on three machines, each with a NVIDIA V100 GPU card with 32GB memory.

BERT and ALBERT's tokenizers are reused to tokenize words into word pieces \cite{kudo-richardson-2018-sentencepiece}. For direct comparing with BERT and ALBERT's event representation learning, we also reuse an existing 100-dimension word embedding file\footnote{\url{http://nlp.stanford.edu/data/glove.6B.zip}} with 400K words (covers 75.76\% of words in the event set) pre-trained by GloVe \cite{pennington2014glove} and set it to be tunable. 

\subsection{Prediction Results and Discussions}\label{subsec:res_baselines}

\begin{table*} \footnotesize
    \centering
    \begin{tabular}{l|rrr|rrr|rrr}
    \hline
 & \multicolumn{3}{c|}{2-category} &  \multicolumn{3}{c|}{3-category} &  \multicolumn{3}{c}{5-category} \\
 & sp500 & dow & nasdaq & sp500 & dow & nasdaq & sp500 & dow & nasdaq \\
 \hline\hline
wordHAN \cite{hu.wsdm2018}  & 79.0  & 73.8  & 71.3  & 67.7  & 64.1  & 63.6  & 56.9  & 52.8  & 47.7 \\
ECK + event \cite{ding-etal-2019-event-representation}  & 69.7  & 68.2  & 65.1  & 62.1  & 59.5  & 56.9  & 54.9  & 47.7  & 44.6 \\
HATS \cite{Kim2019HATSAH}  & 68.7  & 67.7  & 65.1  & 58.5  & 54.9  & 51.8  & 51.8  & 47.2  & 45.6 \\
docBERT \cite{DBLP:journals/corr/abs-1904-08398}  & 85.1  & 80.0  & 72.3  & 71.3  & 69.7  & 64.1  & 59.0  & 55.4  & 49.7 \\
\hline  
GloVe + eventHAN (ours)  & 89.7  & 86.2  & 76.4  & 77.4  & 74.4  & 67.2  & 61.5  & 58.5  & 52.8 \\
BERT + eventHAN (ours)  & 91.3  & 88.2  & 79.0  & 80.0  & 76.4  & 70.8  & 66.2  & 60.5  & 54.9 \\
ALBERT + eventHAN (ours)  & \textbf{93.3}  & \textbf{90.3}  & \textbf{83.1}  & \textbf{84.1}  & \textbf{79.0}  & \textbf{74.4}  & \textbf{70.3}  & \textbf{64.6}  & \textbf{57.9} \\

\hline
    \end{tabular}
    \caption{2/3/5-category accuracies (\%) under 2013 test set.}
    \label{tab:results_235c_3indices}
\end{table*}

We report three variants of our framework basing on event-level HAN: GloVe-based of word2vec style, BERT and ALBERT-based of pretraining + tuning style. ALBERT \cite{Lan2020ALBERT:} is a lite BERT which significantly outperform BERT through three updates: factorized embedding parametrization, cross-layer parameter sharing and inter-sentence coherence loss.

We compare our framework variants with four strong baselines, as shown in Table \ref{tab:results_235c_3indices}. The first is wordHAN \cite{hu.wsdm2018} which utilized words in news for 3-category individual stock movement prediction in Chinese market. Their original accuracies were reported to be around 40\% to 50\%. We reimplement their network and retrain it using our datasets for US market prediction. The second is ECK+event \cite{ding-etal-2019-event-representation}, i.e., external commonsense knowledge (ECK) enhanced event representative learning. Since only 2-category results were reported in \cite{ding-etal-2019-event-representation}, we reuse their code\footnote{\url{https://github.com/MagiaSN/CommonsenseERL\_EMNLP\_2019}} and retrain it for 3-category and 5-category predictions. Note that, even using exactly the same news data, the number of events used in \cite{ding-etal-2019-event-representation} is only 366K. When we replace their event set with ours and retrain their model, we obtain an averagely 4.2\% absolute improvements of the 9 tasks. This reflects the importance of mining large-scale and coreference-free events (Figure \ref{figure:event_extraction_pipeline}).

The third is HATS, a hierarchical graph attention network \cite{Kim2019HATSAH} on historical stock prices and company-relations. Their original averaged accuracies for 3-category S\&P 500 index prediction was under 40\%. We reuse their code\footnote{\url{https://github.com/dmis-lab/hats}} and retrain it by enriching relations with our events which also include named entity relations. This updating brings more than 10\% improvements (Table \ref{tab:results_235c_3indices}). The fourth is a document-classification oriented BERT \cite{DBLP:journals/corr/abs-1904-08398}. Since their original idea and code are only for single-document classification yet we have hundreds of news documents per day, we modify their code\footnote{\url{https://github.com/castorini/hedwig}} to include an additional recurrent+attention mechanism (same with the usage in our HAN, Figure \ref{figure:han}) so that document vectors represented by docBERT are further processed to yield a final classification result.

We conclude the major results. First, ALBERT+eventHAN performs significantly better ($p<0.01$) than its GloVe+eventHAN counterpart, with an absolute improvement of averagely 5.9\%. This observation also aligns with the recent success of ``pretraining+tuning'' architecture in numerous NLP tasks \cite{2020arXiv200308271Q}. Second, ALBERT+eventHAN performs significantly better ($p<0.01$) than docBERT, with an absolute improvement of averagely 10.0\%. The improvements come from two folds, ALBERT itself performing significantly better than BERT and the additionally appended eventHAN. To remove the impacts from ALBERT, we compare BERT+eventHAN with docBERT: BERT+eventHAN is still significantly better ($p<0.01$) than docBERT reflecting the effectiveness of eventHAN (+6.7\% averagely).

Among the four baselines, docBERT performs the best, showing the BERT-style models' strength. Even respectively enriched with external commonsense knowledge and wikidata, ECK+event and HATS did not outperform wordHAN. Despite this, graphical neural network is a promising direction and both its theory and applications are developing in a fast way. Enriching HATS with large-scale textual data is supposed to be a valuable direction.

In addition, in Table \ref{tab:results_235c_3indices}, we observe that the difficulties of predicting S\&P 500, Dow Jones and Nasdaq are improving reflected by their absolute accuracies. We found that in the news dataset, the amount of news mentioning Dow Jones and Nasdaq is respectively only 80\% and 50\% of that for S\&P 500 and the IT companies included in Nasdaq index changes the most frequently. 

\subsection{Is It Really Correct of Using BERT Here?}

Are BERT-style pretraining models really suitable for stock movement prediction? One serious question is that, what if the data used for pretraining BERT/ALBERT are from the same period of the test sets (year of 2013 in this paper and in our references) and already include hints of the market movements? Even the usage of GloVe word2vec is doubtful, so does the external commonsense knowledge. That is, to make the prediction align with real-world applications, no ``future'' related data should be included for stock movement predicting, regardless if they are news, external resources or wikidata: their creation timestamps matter. 

In order to answer this question, we need to set our test set to be \emph{after} the releasing of these pretraining models or external data. The ALBERT model was the latest, released at 2019/12/30 from their web page. We thus construct another test set of covering these three indices of the first four months (83 market-opening days) of 2020, which means no external data are in this period. We set the new development period (127 days) to be from 2019/07/01 to 2019/12/31. The remaining 2006/10/20 to 2019/06/30 with 3,194 days are taken as the new training set.

Moreover, we further collected the Bloomberg and Reuters news data of from Nov. 2013 to the end of Apr. 2020. Then, we perform the same event extraction pipeline and unify the events to construct the new training, validating and testing sets. Major statistical information and 5-category results are listed in Table \ref{tab:2020_data} and Table \ref{tab:results_5c_3indices_2020}, respectively. 

\begin{table}[] \footnotesize
    \centering
    \begin{tabular}{l|rrr}
    \hline
 & Train & Validation & Test \\
\hline\hline
\# samples & 3,194 & 127 & 83 \\
\# news/sample & 335.9 & 491.9 & 501.4 \\
\# events/sample & 17,031.5 & 18,930.5 & 19,538.2 \\
\# words/event & 8.0 & 7.9 & 7.9 \\
\hline
    \end{tabular}
    \caption{Statistical information for updated three sets.}
    \label{tab:2020_data}
\end{table}

\begin{table} \footnotesize
    \centering
    \begin{tabular}{l|rrr}
    \hline
 & sp500 & dow & nas. \\
 \hline\hline
wordHAN \cite{hu.wsdm2018}  & 49.4  & 45.8  & 42.2 \\
ECK + event \cite{ding-etal-2019-event-representation}  & 47.0  & 43.4  & 39.8 \\
HATS \cite{Kim2019HATSAH}  & 43.4  & 41.0  & 38.6 \\
docBERT \cite{DBLP:journals/corr/abs-1904-08398}  & 55.4  & 51.8  & 47.0 \\
\hline
GloVe + eventHAN (ours)  & 57.8  & 55.4  & 50.6 \\
BERT + eventHAN (ours)  & 62.7  & 59.0  & 53.0 \\
ALBERT + eventHAN (ours)  & \textbf{65.1} & \textbf{61.4} & \textbf{55.4}\\
\hline
    \end{tabular}
    \caption{5-cat. accuracies (\%) under 2020 test set.}
    \label{tab:results_5c_3indices_2020}
\end{table}

There is an increasing of news and events per day during recent years, comparing Table \ref{tab:num_news_events} and Table \ref{tab:2020_data}. These also bring longer event sequences and bigger challenge of employ BERT-style pretraining models for fine-tuning. For simplicity, we only report the most difficult 5-category prediction. Our proposed approaches still significantly outperforms ($p<0.01$) the four baselines. The improvements of the systems are comparable to that listed in Table \ref{tab:results_235c_3indices}, a averagely +9.2\% absolute improvements of ALBERT+eventHAN compared with the best baseline of docBERT. However, generally the accuracies drop in the 2020 test set, averagely -3.6\%, compared with the 2013 test set. The reasons are multi-fold: the indices' movements of year 2020 are extremely serious (with tripled standard derivations compared with its former year) and thus less predictable due to the worldly influenced COVID-19 virus, and no 2020-year data are taken into consideration for retraining BERT, ALBERT or other knowledge datasets used in the baseline systems.   

\subsection{Application to Individual Stocks}

Even our major target is to understand systematic behaviours in stock market taking predicting of indices as our task, we are still wondering how good our proposed methods at learning individual stocks. Following \cite{ding-etal-2015-ijcai,ding-etal-2016-knowledge}, we select ten companies, including Google, Microsoft, Apple, Visa, Nike, Boeing, Wal-Mart, Starbucks, Symantec, and Avon Products. We pick the 2020's first four months as our test set and the other configurations follow Table \ref{tab:2020_data}. We report their averaged 5-category accuracies of the four baselines and our three system variants. Under the same order of Table \ref{tab:results_5c_3indices_2020}, the four baselines achieve accuracies of 43.2\%, 40.1\%, 38.7\%, and 45.9\%. Our three system variants achieve accuracies of 50.6\%, 53.8\% and 55.7\%, all are significantly better ($p<0.01$) than the four baselines. Generally, the prediction of individual stocks in our experiments are more difficult compared with index prediction. One reason is the amount of daily news reflecting individual companies are quite limited and the other reason is that individual companies have rather long-term developing strategies which require larger window size of historical days, months or even years, which further brings computational difficulties. 

\subsection{Market Trading Simulation}

We finally evaluate our framework by simulating (back-testing) the stock trading during the 83 market-opening days of 2020. We conduct the trading in daily frequency. Initially, we suppose we hold these 3 indices and 10 stocks at equal amounts. At the beginning of a trading date, our model will give each index and stock a score based on the probability of the five movement categories: we respectively short/long 100\% of current amount of indices in case of DOWN-/UP+, and 50\% in case of DOWN/UP and keep unchanging in case of PRESERVE. It is possible that we sell out all the indices or stocks, then hold the cash, and buy them (top-$5$) back when there is predicted a UP or UP+ the next day. We take a transaction cost of 0.3\% for each trading. We use the annualized return as the metric, which equals to the cumulative profit per year. Following the order in Table \ref{tab:results_5c_3indices_2020}, the four baseline systems respectively yield 61.0\%, 57.2\%, 52.1\%, and 71.9\% annualized return. Our three variants respectively achieved 85.2\%, 88.4\% and 93.2\% annualized returns. This simulation results additionally verify that our proposed approach is more robust compared with baseline systems.

\subsection{Explanation of Systematic Behaviours}

\begin{table} [ht!] \footnotesize
    \centering
    \begin{tabular}{c|l}
    \hline
$c$  &  $weight$: Events\\ 
\hline\hline
D-  &  \textbf{0.52}: the standard \& poor 's 500 index .spx  \\
    &          \ \ \underline{lost} 27.75 points.\\ 
  &  \textbf{0.42}: the dow jones industrial average .dji \\
  & \ \ \underline{dropped} 216.40 points.\\ 
  &    0.02: the fed \underline{stop buying} bonds.\\ 
  &    0.01: the decline marks \\
  &  \ \ \underline{the biggest percentage drop}.\\ 
  &    0.01: the nasdaq composite index .ixic \\
  &  \ \ \underline{fell} 45.57 points.\\ 
\hline 
D  &     \textbf{0.49}: the company took the \underline{pop}.\\ 
  &     \textbf{0.43}: the world 's \underline{second-biggest economy} \\
   &         \ \ was still \underline{losing} \underline{momentum}.\\ 
  &     0.03: us 're going to \underline{get out of} earnings.\\ 
  &     0.02: the nasdaq composite index clung to u.s. \\
   &           \ \ blue-chip stocks gains.\\ 
  &     0.01: chinese activity \underline{slowed} to an 11-month \textit{low}.\\ 
\hline
P     &  \textbf{0.39}: company has \underline{given over} the last month.\\ 
     &  \textbf{0.23}: the horizon are the fed meeting.\\ 
     &  \textbf{0.21}: the last month or so said michael sheldon.\\ 
     &  0.09: i think verizon.\\ 
     &  0.03: comments are different than he.\\ 
\hline 
U  &     \textbf{0.56}: analysts \underline{cautioned} on price \underline{swings}.\\ 
  &     0.25: brent crude oil dipped briefly.\\ 
  &     0.14: investors were also eying friday 's  \\ 
    &           \ \ monthly u.s. non-farm payrolls data.\\ 
  &     0.02: people think the fed.\\ 
  &     0.01: the two have promising \\ 
  &  \ \ \textit{cancer immunotherpaies}.\\ 
\hline
U+  &     \textbf{0.57}: stocks fared \textit{better} in europe.\\ 
  &     0.20: investors looked to \textit{take advantage of}  \\ 
     &             \ \ the previous session 's sharp sell-off.\\ 
  &     0.10: it further \underline{undermined} confidence in the \\
  &  \ \ state of recovery.\\ 
  &     0.03: \underline{divergent} views said jack ablin.\\ 
  &     0.03: i do \textit{believe} there.\\ \hline
    \end{tabular}
    \caption{Event examples ranked by attention mechanisms for 5-category classification. D=Down, P=Preserve, and U=UP.}
    \label{table_c4_7_5c_eventlist}
\end{table}
We finally investigate the explanation ability of our model in terms of event sequences. Table \ref{table_c4_7_5c_eventlist} lists the top-5 events ranked by attention mechanisms for 5-category S\&P500 classification of 5 days. All these 5 days are correctly predicted by our ALBERT+eventHAN model. Negative words, such as \emph{lost}, \emph{dropped}, \emph{stop buying}, \emph{the biggest percentage drop}, and \emph{fell}, appear in all the top-5 related events of the DOWN- category. This reflects that the \emph{strength} of the news does be taken into consideration by the investors and the market. In the DOWN category, most events include negative words, such as \emph{losing momentum}, \emph{get out of}, and  \emph{slowed .. low}. Comparing these five events with DOWN-'s top-5 events, we have a sense that DOWN-'s events contain more strong negative words with specific numbers such as ``lost 27.75 points'', ``dropped 216.40 points'', ``the biggest percentage drop'' and ``fell 45.57 points''. These also provide us an evidence of the \emph{impact} of the financial events to the final stock movement prediction. 

The top ranked events in the PRESERVE category are rather more neutral without positive or negative words. In the UP and UP+ categories, most events contain positive words, such as \emph{better} and \emph{take advantage of}. Note that there are also neutral or slightly negative words used, such as \emph{cautioned .. swings}, \emph{dipped briefly}, \emph{undermined}, and \emph{divergent}. These reflect that the latent linkage between the polarities of the financial events and S\&P 500 index movements is more than linear. That is, daily news contains both positive and negative news and it is important for us to model the impact of each news and their final combined contribution to the final movement. These reflect the importance of the event-driven hierarchical attention network with pretrained contextualized language models.

\section{Related Work}

Employing event representations for stock movement prediction has been proposed in \cite{ding-etal-2015-ijcai,ding-etal-2016-knowledge,ding-etal-2019-event-representation} for index and individual stock prediction. For example, external commonsense knowledge, such as the intents and emotions of the event participants, was utilized in \cite{ding-etal-2019-event-representation} to enhance event representation learning. We follow the usage of events dynamically extracted from financial news. The differences are that we additionally perform a neural coreference resolution module to keep the events being independent and we did not perform any manual filtering (refer to Figure \ref{figure:event_extraction_pipeline} and Table \ref{tab:results_235c_3indices}).

In addition to event representation learning followed by shallow neural networks, deep HANs \cite{yang-etal-2016-hierarchical,hu.wsdm2018} that embeds various granularities of market document-style information are another tendency. Hierarchical graph attention networks \cite{Kim2019HATSAH} made use of existing corporate relational data from Wikidata. Examples of these relational data are alike triplets of \emph{[Apple, Founded by, Steve jobs]}, which align with the commonsense knowledge used in \cite{ding-etal-2019-event-representation}. Graph neural networks by incorporating company knowledge graphs which express inter-market and inter-company relations were proposed in \cite{Matsunaga2019ExploringGN} for Japanese stock market prediction. In stead of using existing relational data, our pipeline keeps updating itself by extracting updated events from lastly published news. In our eventHANs, the final predictions are \emph{explainable} in terms of which events expressed in which news published in which day (Table \ref{table_c4_7_5c_eventlist}).

On the other hand, pretrained contextualized language models, such ELMo \cite{peters-etal-2018-deep}, BERT \cite{devlin-etal-2019-bert}, GPT \cite{Radford2018ImprovingLU} and their consequent variants \cite{2020arXiv200308271Q} are leading the research in numerous NLP applications. However, most existing pretraining models can only take limited length of sequences (such as 512 tokens) as inputs while there are thousands of news containing million-level words appearing everyday. Each news expresses limited information and investors are required to accumulate news together to predict their influence to the future market. In this paper, we combine three elements together: \emph{event streams} represented by \emph{BERT/ALBERT} and their integration in \emph{HANs} to capture super long event sequences for better stock movement prediction. The strengths of our pipeline include: (1) significantly large-scale syntactically independent event sequences are extracted, (2) extremely long event sequences are leveraged, and (3) explainable predictions with accuracies significantly better than state-of-the-art baselines are achieved.

\section{Conclusion}

In this paper, we investigate answers to the question if textual information such as financial events can qualitatively and quantitatively influence the stock market's movements. Our contributions to this field include: a neural co-reference enhanced OpenIE pipeline for event extraction from financial news, BERT/ALBERT enhanced event representations, an event-enhanced HAN that utilizes event, news and temporal attentions, and significantly better accuracies and simulated annualized returns than four state-of-the-art baselines on 3 indices and 10 stocks. We observe that quantitative prediction is feasible in a sense that the strength or importance of news is successfully understood and absorbed by the market. This aligns with the efficient-market hypothesis that asset prices reflect all available information\footnote{Eugene Fama won Nobel Memorial Prize in Economics (2013) for collectively expanding the understanding of asset prices.} \cite{doi:10.1111/j.1540-6261.1970.tb00518.x}. 

For sure, there are a lot of future work: (1) modelling and predicting of global scale markets such as Nikkei 225, TOPIX in Japan, HSI index in Hong Kong, and (2) integrate textual information with financial asset price models, such as Capital Asset Pricing Model (CAPM) \cite{10.2307/2977928}, Arbitrage pricing theory (APT) \cite{apt1976}, and multi-factor models \cite{NBERw20592} will be one interesting direction that combines NLP techniques and finance theory through deep neural networks for a same target of future asset pricing: investors read both textual finance information and digital finance indicators.

\bibliography{emnlp2020}
\bibliographystyle{acl_natbib}

\end{document}